\documentclass[twocolumn,showpacs,prl]{revtex4}
\usepackage{epsfig,amsfonts,amssymb}
\newcommand{\eqref}[1]{(\ref{#1})}

\newcommand{\sumtwo}[2]%
{\mathop{\sum_{#1}}_{#2}}

\newcommand{\fp}[2]{\frac{\partial #1}{\partial #2}}

\newcommand{\fd}[2]{\frac{d#1}{d#2}}

\newcommand{\fdt}[1]{\fd{#1}{t}}

\newcommand{\qedm}{\rule{1.5mm}{3mm}}

\newcommand{\tr}[1]{\overline{#1}}
\newcommand{\bp}{{\bf p}}
\newcommand{\br}{{\bf r}}
\newcommand{\bF}{{\bf E}}
\newcommand{\bse}{{\bf e}}

\newcommand{\calD}{{\cal D}}
\newcommand{\calB}{\mathcal{B}}
\newcommand{\calW}{\mathcal{W}}
\newcommand{\calX}{\mathcal{X}}
\newcommand{\map}{\mathcal{T}}
\newcommand{\tLa}{\widetilde{\Lambda}}
\newcommand{\bR}{\mathbf{R}}
\newcommand{\bP}{\mathbf{P}}
\newcommand{\bbR}{\mathbb{R}}
\newcommand{\ep}{\varepsilon}
\newcommand{\tPT}{\tilde{P}^{(T)}}
\newcommand{\Trel}{T_{\rm rel}}
\newcommand{\Twait}{T_{\rm wait}}
\newcommand{\Tper}{T_{\rm per}}

\begin{document}
\title{Two theorems that relate discrete stochastic processes to  microscopic mechanics}
\author{Hal Tasaki${}^{\dagger}$}
\affiliation {Department of Physics, Gakushuin University,
Mejiro, Toshima-ku, Tokyo 171-8588, Japan}

\date{\today}

\vspace{5in}

\begin{abstract}
Starting from a classical mechanics  of a ``colloid particle''  and $N$ ``water molecules'',  we study effective stochastic dynamics of the particle which jumps between deep potential wells.
We  prove that the effective transition probability satisfies (local) detailed balance condition.
This enables us to rigorously determine precise form of the transition probability when  barrier potentials have certain regularity and symmetry.
\end{abstract}

\maketitle

Construction of statistical mechanics that applies to systems far from equilibrium is a major challenge in theoretical physics (see \cite{SST} and references therein).
It is expected that simple stochastic processes with discrete state spaces may be studied to elucidate various nonequilibrium phenomena and universal features of nonequilibrium states \cite{DLG}.
To define a physically meaningful stochastic processes, however, is not a straightforward task.

There has been a consensus that the {\em detailed balance condition}\/ is both necessary and sufficient for recovering physically realistic dynamics near and at equilibrium.
But there has been no such guiding principles for dynamics far from equilibrium.
Moreover it was pointed out recently that the nature of steady states depend drastically  on the choice of stochastic dynamics in driven many-particle systems \cite{Ruledepedence}.
With such a strong rule dependence one can hardly extract physically meaningful information from these stochastic models.
It now seems necessary to  reexamine what are ``physically realistic'' choices of stochastic dynamics.
In doing that, one should avoid employing any scheme (including Langevin equation) whose validity far from equilibrium has not been established.

Here we present such a reexamination staring from microscopic  mechanics.
We consider, as probably the simplest nontrivial example, a classical mechanical system consisting of a ``colloid particle''  and $N$ ``water molecules''.
We assume that water is in thermal equilibrium while the particle is trapped in one of many deep potential wells.
We focus on the effective stochastic dynamics of the particle which jumps from a potential well to another.
We first show that  the transition probability satisfies the {\em detailed balance condition}\/.
The result extends to the situation where the particle (not molecules) is driven by an external field, thus establishing the  {\em local detailed balance condition}\/ \cite{DLG}.
By using these general results, we can essentially determine the transition probability when the potential barriers has certain regularity and symmetry.
The resulting form \eqref{e:Pfinal}, in which the barrier height determines the transition probability, has been well-known \cite{classical}.
Since this result applies to driven systems as well, it provides a starting point for reexamining various features exhibited by nonequilibrium systems.

\paragraph*{Setting:}
We consider a system of  a single ``colloid particle'' (which we call a particle) and $N$ ``water molecules'' (which we call molecules)  in a finite $d$-dimensional periodic box  $\Lambda\subset\bbR^d$.
We denote by $\bR\in\Lambda$ and $\bP\in\bbR^d$ the coordinate and momentum, respectively, of the particle, and by $\br_i\in\Lambda$ and $\bp_i\in\bbR^d$ the coordinate and momentum, respectively, of the $i$-th molecule where $i=1,2,\ldots,N$.
We denote the state of the system collectively as
$\Gamma=(\bR,\bP;\br_1,\ldots,\br_N;\bp_1,\ldots,\bp_N)$,
and the corresponding Lesbegue measure as
$d\Gamma=d^d\bR\,d^d\bP\prod_{i=1}^Nd^d\br_i\,d^d\bp_i$.
The Hamiltonian is
\begin{equation}
H(\Gamma)=\frac{|\bP|^2}{2M}+V(\bR)+\sum_{i=1}^N\frac{|\bp_i|^2}{2m_i}+U(\bR;\br_1,\ldots,\br_N),
\label{e:H}
\end{equation}
where $V(\cdot)$ represents the external  force acting on the particle, and $U(\cdot)$ represents the interaction between the particle and molecules, the interaction between molecules, and the external force acting on molecules.
We denote by $\map_t(\cdot)$ the time evolution map of the Hamiltonian dynamics determined by \eqref{e:H}.

\begin{figure}
\begin{center}
\epsfig{file=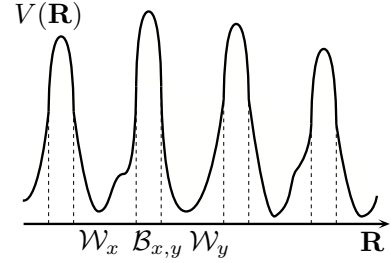,width=5cm}
\end{center}
\caption{
A typical potential $V(\bR)$ and wells.
Here the potential at the barriers has regularity and symmetry stated below.
}
\label{f:pot}
\end{figure}

Let $\calX$ be a finite set.
With each $x\in\calX$, we associate a region $\calW_x\subset\Lambda$ so that $\calW_x\cap\calW_y=\emptyset$ whenever $x\ne y$.
Each $\calW_x$ corresponds to a potential well in which the particle may be trapped.
We thus assume that the potential  $V(\cdot)$ takes relatively small values inside each $\calW_x$, and becomes very large near and in boundary regions separating  different wells (Fig.~\ref{f:pot}).

Suppose that molecules are in (near) equilibrium, and the particle is in a well $\calW_x$.
After a certain amount of time, which we call the relaxation time $\Trel$, the system is expected to reach a quasi-equilibrium where the particle is trapped within $\calW_x$.
This state can naturally be modeled by the restricted canonical distribution
\begin{equation}
P_x(\Gamma)=\frac{1}{Z_x}\,\chi_x(\Gamma)\,e^{-\beta\,H(\Gamma)},
\label{e:P0i}
\end{equation}
where
\begin{equation}
\chi_x((\bR,\bP;\br_1,\ldots,\br_N;\bp_1,\ldots,\bp_N))=
\cases{
1,&if $\bR\in\calW_x$
\cr
0,&if $\bR\notin\calW_x$
}
\label{e:chii}
\end{equation}
is the characteristic function for the event that the particle is in $\calW_x$.
The restricted partition function is
\begin{equation}
Z_x=\int d\Gamma\,\chi_x(\Gamma)\,e^{-\beta\,H(\Gamma)}.
\label{e:Zi}
\end{equation}

\paragraph*{Detailed balance condition:}
Let us study the probability $P^{(T)}_{x\to y}$ that the particle moves from a well $\calW_x$ to another well $\calW_y$ after a fixed time $T>0$.
To do this, we sample initial condition according to \eqref{e:P0i},  let the whole system evolve for time interval $T$ according to the  Hamiltonian dynamics, and then ask whether the particle is  in $\calW_y$.
The resulting probability is
\begin{equation}
P^{(T)}_{x\to y}=\int d\Gamma\,\chi_y(\map_T(\Gamma))\,P_x(\Gamma).
\label{e:Pij}
\end{equation}

Let $\Twait$ be the typical waiting time that the particle spends in a single well.
By letting the potential at barriers high enough we can assume $\Trel\ll\Twait$.
Then by taking $T$ which satisfies $\Trel\ll T\ll\Twait$, we can interpret $P^{(T)}_{x\to y}\ (\ll1)$ as the transition probability from a quasi-equilibrium state where the particle is in $\calW_x$ to another quasi-equilibrium state where it is in $\calW_y$.
We thus end up with an effective Markov process   (with discrete state space and discrete time) of the particle which jumps from a well to another according to the transition probability $P^{(T)}_{x\to y}$ \footnote{
We here do not attempt to justify the existence of the effective Markov process (although, physically speaking, it is obvious that there is a well defined Markov process if $\Trel\ll T\ll\Twait$).
In fact the nature of the effective process depends crucially on details of the model, and such a justification is possible only when one makes various concrete (and sometimes technical) assumptions about the potential and the interaction.
Interestingly {\em all}\/ the results in the present paper hold rigorously, independently to the existence of an effective Markov process.
Indeed all the results are true (but just meaningless) for a model with $U(\cdot)=0$, which is nothing but a deterministic mechanics (with a stochastic initial condition) of a single particle.
}.


Of course there is no hope of evaluating the transition probability  \eqref{e:Pij}  exactly as it involves full Hamiltonian time evolution of the system with a huge number of degrees of freedom.
But, by a modification of the methods in \cite{JMN}, we can easily show the following symmetry.

{\em Theorem 1---}\/
For any $x,y\in\calX$ and  $T>0$, one has
\begin{equation}
Z_x\,P^{(T)}_{x\to y}=Z_y\,P^{(T)}_{y\to x}.
\label{e:DB0}
\end{equation}

If we define the (restricted) free energy $F_x$ by $Z_x=e^{-\beta\,F_x}$, \eqref{e:DB0} becomes
\begin{equation}
e^{-\beta\,F_x}\,P^{(T)}_{x\to y}=e^{-\beta\,F_y}\,P^{(T)}_{y\to x},
\label{e:DB1}
\end{equation}
which is nothing but the detailed balance condition.

{\em Proof---}\/
Let us change the integration variable in \eqref{e:Pij} according to $\Gamma'=\map_T(\Gamma)$.
By using the Liouville theorem $d\Gamma=d\Gamma'$ and the energy conservation law $H(\Gamma)=H(\Gamma')$, \eqref{e:Pij} becomes
\begin{eqnarray}
P^{(T)}_{x\to y}=\frac{1}{Z_x}\int d\Gamma'\,\chi_x(\map_{-T}(\Gamma'))\,\chi_y(\Gamma')\,e^{-\beta\,H(\Gamma')}.
\label{e:Pij2}
\end{eqnarray}
We define the time reversal of a state $\Gamma$ by
$\tr{\Gamma}=(\bR,-\bP;\br_1,\ldots,\br_N;-\bp_1,\ldots,-\bp_N)$,
where we have reversed all the momenta.
Note that for any $\Gamma$ and $x$, we have $\chi_x(\Gamma)=\chi_x(\tr{\Gamma})$.
By using the time reversal symmetry we see that
$\tr{\map_{-T}(\Gamma')}=\map_T(\tr{\Gamma'})$.
By also noting that $H(\Gamma')=H(\tr{\Gamma'})$ and $d\Gamma'=d\tr{\Gamma'}$, we have
\begin{eqnarray}
&&P^{(T)}_{x\to y}=\frac{1}{Z_x}\int d\tr{\Gamma'}\,\chi_x(\map_{T}(\tr{\Gamma'}))\,\chi_y(\tr{\Gamma'})\,e^{-\beta\,H(\tr{\Gamma'})}
\nonumber\\
&&=\frac{1}{Z_x}\int d\Gamma\,\chi_y(\Gamma)\,\chi_x(\map_T(\Gamma))\,e^{-\beta\,H(\Gamma)}
=\frac{Z_y}{Z_x}P^{(T)}_{y\to x},
\label{e:Pij3}
\end{eqnarray}
where we have renamed the variable $\tr{\Gamma'}$ as $\Gamma$ to get the second line \footnote{
It is obvious that the present theorem and proof extend to much more general situations.
What one needs are disjoint subsets $\calD_1,\ldots,\calD_n$ of the whole phase space, where the system has to  overcome a high energy barrier to get out of each $\calD_i$.
Then for a given Hamiltonian $H$, we define the initial quasi-equilibrium distribution by generalizing \eqref{e:P0i} as $P_i(\Gamma)=(Z_i)^{-1}\chi[\Gamma\in\calD_i]\,e^{-\beta\,H(\Gamma)}$.
The transition probability \eqref{e:Pij} is also generalized as $P^{(T)}_{i\to j}=\int d\Gamma\,\chi[\map_T(\Gamma)\in\calD_j]\,P_i(\Gamma)$.
Then one can repeat the same argument to prove the detailed balance condition.
In this manner, we can treat essentially any many-body systems.
}.~\qedm


\paragraph*{Driven systems:}
It is crucial that we have derived the condition \eqref{e:DB1} without referring to the global equilibrium of the system.
All we needed was the Hamiltonian mechanics in which the particle moves relatively short distance.
This allows us to extend the result to driven systems, which never settle to global equilibrium.

We now assume that, in addition to the force described by the potential $V(\bR)$, a constant driving force $\bF$ acts on the particle.
The resulting mechanics {\em cannot}\/ be described by a Hamiltonian mechanics (since we impose periodic boundary conditions), but by Newtonian equations as
\begin{equation}
\fdt{\bR}=\fp{H}{\bP},\ 
\fdt{\bP}=-\fp{H}{\bR}+\bF,\ 
\fdt{\br_i}=\fp{H}{\bp_i},\ 
\fdt{\bp_i}=-\fp{H}{\br_i},
\label{e:EMF}
\end{equation}
for $i=1,\ldots,N$.

Our derivation of \eqref{e:DB1} is restricted to a Hamiltonian system, and does not apply to the Newtonian system \eqref{e:EMF} as it is.
We argue, however, that this Newtonian system can be replaced by a Hamiltonian system as long as the short-time behavior is concerned.

Suppose that the system evolves according to \eqref{e:EMF} starting from a distribution in which the particle is in $\calW_x$.
There is a characteristic time $\Tper$ in which the particle travels around the whole space and sees for the first time that $\Lambda$ is periodic.
Since $\Tper$ grows indefinitely as the linear size of the system increases, we can safely assume that $T\ll \Tper$ .
Then we take a smaller box $\tLa\subset\Lambda$ which includes $\calW_x$ and is large enough for the particle to remain always in $\tLa$ during the time interval $T$.
We define the new Hamiltonian \footnote{
If  necessary we use periodic boundary conditions to redefine $\bR$ so that it has no jumps within $\tLa$.
}
\begin{equation}
H_{\bF}=H-\tilde{\bF}(\bR)\cdot\bR.
\label{e:HF}
\end{equation}
where $\tilde{\bF}(\bR)=\bF$ if $\bR\in\tLa$ and $\tilde{\bF}(\bR)=0$ otherwise. 
As long as the particle stays within $\tLa$, the equation of motion determined by \eqref{e:HF} is exactly the same as \eqref{e:EMF}.

We can then define the quasi-equilibrium distribution \eqref{e:P0i} using the new Hamiltonian \eqref{e:HF}.
Since the distribution \eqref{e:P0i} and the partition function \eqref{e:Zi} only involves $\bR$ in $\calW_x\subset\tLa$, the sharp cutoff at the boundary of $\tLa$ in \eqref{e:HF} does not affect \eqref{e:P0i} or \eqref{e:Zi}.
Note that the restricted partition function $Z_x^{(\bF)}$ (and hence the restricted free energy $F_x^{(\bF)}$) may depend on the choice of $\tLa$, 
but the difference $F_x^{(\bF)}-F_y^{(\bF)}$ has a definite value when the wells $\calW_x$ and $\calW_y$ are close to each other.
We can then repeat the same argument as before to derive
\begin{equation}
e^{\beta\,(F_y^{(\bF)}-F_x^{(\bF)})}\,P^{(T)}_{x\to y}=P^{(T)}_{y\to x},
\label{e:final2}
\end{equation}
which is known as the {\em local detailed balance condition}\/, and has been used as a fundamental requirement in designing stochastic processes of driven systems \cite{DLG}.

\paragraph*{Determination of the transition probability:}
In order to proceed further and determine the transition probability $P^{(T)}_{x\to y}$, we need to assume that potential at barriers between the wells are identical in their shapes (but not necessarily in their heights) and that each barrier potential looks identical from both the sides.
The potential within the wells is completely arbitrary.

\begin{figure}
\centerline{\epsfig{file=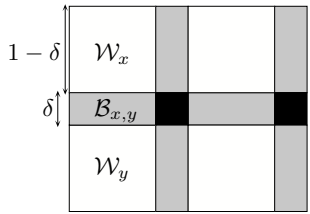,width=4cm}}
\caption[dummy]{
Arrangement of the wells and barriers in two dimensions.
}
\label{f:WB}
\end{figure}

For simplicity we restrict ourselves to the simplest geometry depicted in Figs.~\ref{f:pot} and \ref{f:WB}.
Let $\calX=\{1,2,\ldots,L\}^d\subset\Lambda$  be the unit lattice.
We fix a constant $0<\delta<1$.
For each $x\in\calX$, we define the corresponding well as a $d$-dimensional cube $\calW_x=\{\bR\,|\,|R_j-x_j|<(1-\delta)/2\ \mbox{for $j=1,\ldots,d$}\}$, where we wrote $\bR=(R_1,\ldots,R_d)$ and $x=(x_1,\ldots,x_d)$.
For each pair $x,y\in\calX$ such that $|x-y|=1$, we define the boundary region $\calB_{x,y}=\calB_{y,x}$ as follows.
Let $\bse_j$ be the unit vector in the $j$-th direction.
Then for any $x\in\calX$ and $j=1,\ldots,d$, we set
\begin{eqnarray}
\calB_{x,x+\bse_j}&=&\{\bR\,|\,|R_j-(x_j+\frac{1}{2})|\le\frac{\delta}{2},
\nonumber\\&&
\hspace{1cm}
|R_k-x_k|<\frac{1-\delta}{2}\ \mbox{for $k\ne j$}\}.
\label{e:Bxxe}
\end{eqnarray}
To specify the barrier potential, we define the prototype
\begin{equation}
\calB=\{\bR\,|\,|R_1|\le\frac{\delta}{2},|R_k|<\frac{1-\delta}{2}\ \mbox{for $k\ne1$}\},
\label{e:B}
\end{equation}
of the boundary region, 
and let $v_0(\bR)\ge0$ be an arbitrary potential on $\calB$ with the reflection symmetry $v_0(R_1,R_2,\ldots,R_d)=v_0(-R_1,R_2,\ldots,R_d)$.
This means that the barrier looks identical from both the sides.
For each $x,y\in\calX$ such that $|x-y|=1$, there is a map $f_{x,y}:\calB_{x,y}\to\calB$ which consists of a translation and (if necessary) a rotation.

We state our assumption on the potential $V(\cdot)$.
If $\bR\in\calW_x$ for some $x$, then $V(\bR)$ is completely arbitrary.
If $\bR\in\calB_{x,y}$ for some $x,y\in\calX$, then we set $V(\bR)=v_0(f_{x,y}(\bR))+b_{x,y}$, where $b_{x,y}>0$ is an arbitrary (large) constant which characterizes the height of the barrier at $\calB_{x,y}$.
If $\bR\not\in\calW_x$ and $\bR\not\in\calB_{x,y}$ for any $x,y$, then we set $V(\bR)=\infty$.
The potential $U(\bR;\br_1,\ldots,\br_N)$  is assumed to be invariant under any translation by a lattice vector $x\in\calX$.

We first consider Hamiltonian time evolution with the Hamiltonian \eqref{e:H}.
It is possible to determine the transition probability $P^{(T)}_{x\to y}$ in the present rather general setting provided that $T\ll\Twait$.

{\em Theorem 2---}\/
For any $\ep$ with $0<\ep<1$, there exists (a sufficiently small) $T>0$  such that
\begin{equation}
1-\ep\le \frac{P^{(T)}_{x\to y}}{\rho(T)\,e^{-\beta(b_{x,y}-F_x)}}\le
1+\ep
\label{e:Pxybounds}
\end{equation}
holds for any $x,y\in\calX$ with $|x-y|=1$, where $\rho(T)>0$ is independent of $x,y$.

The theorem determines the transition probability as 
\begin{equation}
P^{(T)}_{x\to y}\simeq\rho(T)\,e^{-\beta(b_{x,y}-F_x)}.
\label{e:Pfinal}
\end{equation}
This form has been known for quite a long time \cite{classical}, but we stress that here a neat and completely rigorous justification is given \footnote{
We note that this is the leading behavior in the Krammers' formula.
It seems very difficult to get rigorous estimates corresponding to the subleading behavior.
}.
The physics behind \eqref{e:Pfinal} is clear; the transition is mainly ruled by the process in which the particle thermally obtains  (free) energy $b_{x,y}-F_x$ to overcome the barrier at $\calB_{x,y}$.
 

{\em Proof---}\/
Consider a time evolution starting from an initial state $\Gamma$ distributed according to the quasi equilibrium distribution $P_x(\Gamma)$ in \eqref{e:P0i} for some $x\in\calX$.
For $T>0$ and $y\in\calX$ such that $|x-y|=1$, let $\tPT_{x\to y}$ be the probability that the particle moves to $\calW_y$ directly from $\calW_x$ within  time $T$.
More precisely  $\tPT_{x\to y}$ is the probability that there are $t_1$, $t_2$, and $t_3$ with $0<t_1<t_2<t_3\le T$ such that $\Gamma(t)\in\calW_x$ for $t\in[0,t_1)$, $\Gamma(t)\in\calB_{x,y}$ for $t\in[t_1,t_2]$, and $\Gamma(t)\in\calW_y$ for $t\in(t_2,t_3)$, where $\Gamma(t)=\map_t(\Gamma)$ denotes the state at time $t$.

When $T$ is sufficiently small compared with $\Twait$, one can assume that (with a probability very close to one) the particle executes at most a single jump from $\calW_x$ to a neighboring well within the time interval $T$.
Then the newly defined probability $\tPT_{x\to y}$ should be identical to $P^{(T)}_{x\to y}$.
We therefore see that for any $\ep>0$  there is $T$ such that
\begin{equation}
1-\frac{\ep}{5}\le\frac{P^{(T)}_{x\to y}}{\tPT_{x\to y}}\le1+\frac{\ep}{5}
\label{e:PP1}
\end{equation}
for any $x,y\in\calB$ with $|x-y|=1$.
Then, since $(Z_x\,P^{(T)}_{x\to y})/(Z_y\,P^{(T)}_{y\to x})=1$ from \eqref{e:DB0}, we get
\begin{equation}
\frac{1-\ep/5}{1+\ep/5}\le
\frac{Z_x\,\tPT_{x\to y}}{Z_y\,\tPT_{y\to x}}
\le\frac{1+\ep/5}{1-\ep/5}.
\label{e:ZPZP}
\end{equation}

Now, fix $x,y\in\calB$ with $|x-y|=1$.
We introduce a modified model obtained by replacing $V(\bR)$ for $\bR\in\calW_y$ by a constant $V(\bR)=b_{x,y}-V_0$, and letting $b_{y,z}=\infty$ for all $z$ neighboring to $y$ except $z=x$.
Note that the probability $\tPT_{x\to y}$ does not change by this replacement since $\tPT_{x\to y}$ is fully determined by the dynamics of the particle within $\calW_x$ and $\calB_{x,y}$.
On the other hand  the probability $\tPT_{y\to x}$ becomes very simple since the potential within $\calW_y$ is constant and  the only allowed jump is that to $\calW_x$.
We can write $\tPT_{y\to x}=\tilde{p}_0(T)$ and $Z_y=Z_0\,e^{-\beta\,(b_{x,y}-V_0)}$, where $\tilde{p}_0(T)$ and $Z_0$ are independent of $x,y$.
By denoting $p_0(T)=\tilde{p}_0(T)\,Z_0\,e^{\beta\,V_0}$, \eqref{e:ZPZP} becomes
\begin{equation}
\frac{1-\ep/5}{1+\ep/5}\le
\frac{Z_x\,\tPT_{x\to y}}{p_0(T)\,e^{-\beta\,b_{x,y}}}
\le\frac{1+\ep/5}{1-\ep/5}.
\label{e:ZPZP2}
\end{equation}
Using \eqref{e:PP1} once agin, we get the bounds
\begin{equation}
\frac{(1-\ep/5)^2}{1+\ep/5}\le
\frac{P^{(T)}_{x\to y}}{p_0(T)\,e^{-\beta(b_{x,y}-F_x)}}
\le\frac{(1+\ep/5)^2}{1-\ep/5},
\label{e:ZPZP3}
\end{equation}
which implies the desired \eqref{e:Pxybounds} if one notes $0<\ep<1$.~\qedm

\paragraph*{Driven systems:}
The proof of \eqref{e:Pxybounds} again extends to a driven system with the non-Hamiltonian time evolution \eqref{e:EMF} since we can always study equivalent Hamiltonian system with the Hamiltonian \eqref{e:HF}.
One must note, however, that the assumed symmetry of the barrier potential is usually destroyed by the applied field.

Simple examples where the symmetry survives are those with $\delta=0$, i.e., the models with infinitesimally thin barriers.
In this case we can prove the same estimate \eqref{e:Pxybounds} (with the prefactor $\rho(T)$ possibly being direction dependent).

Usually it is not an easy task to compute the restricted free energy $F_x^{(\bF)}$ since it involves interactions with multiple molecules.
The calculation becomes easy if we assume that the interaction between the particles and the molecules are so small that the partition function almost factorizes.
We further take the simplest Gaussian potential $V(\bR)=(a_x/2)\,|\br-x|^2+c_x$ for $\bR\in\calW_x$ and assume that there is an external field $\bF=(E,0,\ldots,0)$ with $E\ll a_x$ for all $x$.
Then the transition probabilities can be explicitly (and rigorously) computed as
\begin{eqnarray}
&&P^{(T)}_{x\to x\pm\bse_1}\simeq\rho(T)\,\exp[-\beta(b_{x,x\pm\bse_1}-F_x^{(0)}\mp E/2)],
\nonumber\\
&&P^{(T)}_{x\to y}\simeq\tilde{\rho}(T)\,\exp[-\beta(b_{x,y}-F^{(0)}_x)],
\label{e:PTGauss}
\end{eqnarray}
where $y$ is any neighbor of $x$, not in the first direction.
Here
$F^{(0)}_x\simeq c_x-\{3/(2\beta)\}\,\log(Ma_x)+(3/\beta)\,\log(\beta/2\pi)$
is the restricted free energy of the system without an external field.
The prefactors $\rho(T)$, $\tilde{\rho}(T)$ depend on details of the barrier potential.

We stress that \eqref{e:PTGauss} is the ``physically correct'' form of the transition probability in a model of a particle (or of particles) trapped in potential wells and driven by an external field.
Indeed \eqref{e:PTGauss} has been regarded as a realistic transition probability in ionic conductors, which the driven lattice gas is supposed to model.
Given the strong rule dependence \cite{Ruledepedence}, it seems important to study discrete models for nonequilibrium steady states by using the ``correct'' rule \eqref{e:PTGauss}.

I wish to thank Y.~Inaguma, C. ~Jarzynski, C.~Maes, K.~Neto\v{c}n\'{y}, and S. Sasa for useful discussions.


\begin{thebibliography}{10}
\bibitem[$\dagger$]{email-hal}
Electronic address: 
hal.tasaki@gakushuin.ac.jp

\bibitem{SST}
S. Sasa and H. Tasaki, J. Stat. Phys. {\bf 125}, 125 (2006).

\bibitem{DLG} 
S. Katz, J. L. Lebowitz, and H. Spohn, 
J. Stat. Phys. {\bf 34}, 497 (1984);
B. Schimttmann and R. K. P. Zia, 
{\em Statistical mechanics of driven diffusive systems}\/
(Academic Press, 1995).

\bibitem{Ruledepedence}
R. Lefevere and H. Tasaki,
Phys. Rev. Lett.~{\bf 94}, 200601 (2005).

\bibitem{classical}
H. A. Kramers,
Physica {\bf 7}, 284 (1940);
P. G. Bergmann, J. L. Lebowitz, Phys. Rev.~{\bf 99}, 578 (1955).




\bibitem{JMN} 
C. Jarzynski,
J. Stat. Phys. {\bf 98}, 77 (2000);
C. Maes and K. Neto\v{c}n\'{y}, J. Stat. Phys. {\bf 110}, 269
(2003).



\end{thebibliography}
\end{document}